\documentclass[prd,showpacs,preprintnumbers,nofootinbib,aps]{revtex4}
\usepackage{subeqnarray}
\usepackage{epsfig,amsmath,amssymb}
\usepackage{mathrsfs}
\usepackage[usenames,dvipsnames]{color}
\usepackage[pagebackref=false, colorlinks=false]{hyperref}
\definecolor{redish}{rgb}{0.7,0.2,0.0}  
\definecolor{bluish}{rgb}{0.2,0.5,0.8}
\hypersetup{linkcolor=redish,          
                  citecolor=blue,        
                  filecolor=magenta,      
                  urlcolor=bluish}          
\DeclareFontFamily{U}{rsfs}{}         
\DeclareFontShape{U}{rsfs}{m}{n}{<5> rsfs5 <6><7> rsfs7          %
  <8><9><10><10.95><12><14.4><17.28><20.74><24.88> rsfs10}{}     %
\DeclareMathAlphabet{\mathfs}{U}{rsfs}{m}{n}                     %

\newcommand{\ba}{\nopagebreak[3]\begin{eqnarray}}
\newcommand{\ea}{\end{eqnarray}}
\newcommand{\bii}{\begin{itemize}}
\newcommand{\eii}{\end{itemize}}

\begin{document}

\title{Thermal Stability Of Charged Rotating Quantum Black Holes}
\author{Aloke Kumar Sinha}
\email{akshooghly@gmail.com}
\affiliation{Ramakrishna Mission Vivekananda University, Belur Math 711202,India,\\ Haldia Government College,West Bengal, India}
\author{Parthasarathi Majumdar}
\email{parthasarathi.majumdar@rkmvu.ac.in}
\affiliation{Ramakrishna Mission Vivekananda University, Belur Math 711202, India}
\pacs{04.70.-s, 04.70.Dy}
       
\begin{abstract}

 Criteria for thermal stability of charged rotating black holes of any dimension are derived, for horizon areas that are large relative to the Planck area (in these dimensions). The derivation is based on generic assumptions of quantum geometry based on some results of loop quantum gravity, and equilibrium statistical mechanics of  the Grand Canonical ensemble. There is no explicit use of classical spacetime geometry in this analysis. The only assumption is that  the mass of the black hole is a function of its horizon area, charge and angular momentum.  Our stability criteria are then tested in detail against specific classical black holes in spacetime dimensions 4 and 5, whose metrics provide us with explicit relations for the dependence of the mass on the charge and angular momentum of the black holes. This enables us to predict which of these black holes are expected to be thermally unstable under Hawking radiation.

\end{abstract}
\maketitle

\section{Introduction}

It is well-known from semiclassical analysis that nonextremal, asymptotically flat black holes are thermally unstable  due to decay under Hawking radiation, leading to their specific heat being negative \cite{dav77}. This  interesting fact has motivated the study of thermal stability of black holes, from a perspective that is inspired by a definite proposal for {\it quantum} spacetime (like Loop Quantum Gravity (LQG), \cite{rov,thie}) rather than on semiclassical assumptions. A consistent understanding of the issue of {\it quantum} black hole entropy has been obtained through LQG \cite{abck98,abk00}, where not only has the Bekenstein-Hawking area law been retrieved for macroscopic (astrophysical) black holes, but a whole slew of corrections to it, due to quantum spacetime fluctuations have been derived as well \cite{km98}-\cite{bkm10}, with the leading correction being logarithmic in area with the coefficient $-3/2$.  However, we hasten to add the general disclaimer that {\it our paper is neither on LQG, nor does it use the LQG framework in an essential way .} LQG, if anything, plays only a motivational role in our work. Many of the assumptions of the paper, actually made independently of LQG, are justified on the ground that LQG might provide situations where these assumptions are valid.\\
 
The implications of this quantum perspective, on the thermal stability of black holes from decay due to Hawking radiation, have therefore been an important aspect of black hole thermodynamics beyond semiclassical analysis, and also somewhat beyond the strictly equilibrium configurations that Isolated Horizons represent. Classically a black hole in general relativity is characterized by its' mass ($M$), charge ($Q$) and angular momentum ($J$).  Intuitively, therefore, we expect that thermal behaviour of black holes will depend on all of these parameters. For a given classical metric characterizing a black hole, the mass can be derived explicitly to be a function of the charge and angular momentum. However, the quantum spacetime perspective frees us from having to use classical formulae for this functional dependence of the mass. Instead, the assumption is simply this : the mass is a function of the horizon {\it area}, alongwith the charge and angular momentum. \\

The simplest case of vanishing charge and angular momentum has been investigated longer than a decade ago \cite{cm04} - \cite{cm05-2}. This has been  generalized, via the idea of {\it thermal} holography \cite{pm07}, \cite{pm09}, and the saddle point approximation to evaluate the canonical partition function corresponding to the horizon, retaining Gaussian thermal fluctuations. The consequence is a general criterion of thermal stability as an inequality connecting area derivatives of the mass and the microcanonical entropy. This inequality is nontrivial only when the microcanonical entropy has corrections (of a particular algebraic sign) beyond the area law, as is the case for the loop quantum gravity calculation of the microcanonical entropy \cite{km00}. This body of work has been generalized more recently \cite{mm12} to include black holes with charge. The generalized stability criterion indeed `predicts' the thermal {\it instability} of asymptotically flat Reissner-Nordstrom black holes contrasted with the thermal {\it stability} of anti-de Sitter Reissner-Nordstrom black holes (for a range of cosmological constants). \\

In this paper, this approach is generalized to quantum black holes carrying {\it both} charge and angular momentum. The inclusion of rotation poses challenges in the LQG formulation \cite{ak04}, \cite{kkrxx} - \cite{per16} of isolated horizons. However, the general understanding of non-radiant rotating isolated horizons has parallels in these assays. We do not review this body of work, but realize that the thermal stability behaviour of rotating radiant black holes may be {\it qualitatively different} from that of the non-rotating ones. \\ 

The paper is organized as follows: In section 2, the idea of thermal holography, alongwith the concept of (holographic) mass associated with horizon of a black hole is briefly reviewed and the grand canonical entropy of charged rotating large quantum black hole is determined. In section 3, the criterion for thermal stability of such black holes is determined by using saddle point approximation to evaluate the horizon partition function for Gaussian thermal fluctuations around thermal equilibrium. In the next section, this stability criterion is used to test on various explicit classical black holes, with the objective of predicting their behaviour under decay due to Hawking radiation. The next section contains a brief summary and discussion. We end in section 6 with Appendix, showing various calculations in details.

\vspace{.3 in}
\section{Thermal holography}

 In this section, we present a generalization of the thermal holography for non-rotating electrically charged quantum radiant horizons discussed in \cite{mm12}, to the situation when the horizon has both charge and angular momentum. Such a generalization completes the task set out in \cite{cm04} and \cite{pm07} to include  charge and angular momentum in consideration of thermal stability of the horizon under Hawking radiation. This section is really motivational in character, and admittedly heuristic. It has certain parallels with LQG, but does not really represent LQG in any rigorous sense. The main reason for its inclusion is to show that a logical line of argument may exist in a {\it generic} theory of quantum gravity to go from the tensor product Hilbert space to our stability criteria, even though all steps in this chain of arguments may not be on rigorous mathematical footing.
 
\subsection{Mass Associated With horizon}

Black holes at equilibrium are represented by isolated horizons, which are internal boundaries of spacetime. Hamiltonian evolution of this spacetime gives the first law associated with isolated horizon($b$) and is given as \cite{ak04},
\begin{eqnarray}
\delta E^{t}_{h}=\frac{\kappa^{t}}{8\pi}\delta A_{h}+\Phi^{t}\delta Q_{h}+\Omega^{t}\delta J_{h}
\end{eqnarray}
where, $E^{t}_{h}$ is the energy function associated with  the horizon, $\kappa^{t}$, $\Phi^{t}$ and $\Omega^{t}$ are respectively the surface gravity, electric potential and   angular velocity of  the horizon; $Q_{h}~,~A_{h}$ and $J_{h}$ are respectively the charge, area and angular momentum of  the horizon. The label '$t$' denotes the particular time evolution field ($t^{\mu}$)  associated with the spatial hypersurface chosen. $E^{t}_{h}$ is  assumed here to be a function of $A_{h}$, $Q_{h}$ and $J_{h}$.

 The advantage of the isolated (and also the radiant or {\it dynamical}) horizon description is that one can associate with it a mass $M_h^t$, related to the ADM energy of the spacetime through the relation
\begin{eqnarray}
E_{ADM}^t = M_h^t + E_{rad}^t ~\label{adm}
\end{eqnarray}
where, $E^{t}_{rad}$ is the energy associated with spacetime  between the horizon and asymptopia. An isolated horizon does not require stationarity, and therefore admits $E^t_{rad} \neq 0$, and hence admits a mass defined {\it locally} on the horizon.  

\subsection{Quantum Geometry}

The Hilbert space of  a generic quantum spacetime is given as, $\mathcal{H}=\mathcal{H}_{b}{\otimes}\mathcal{H}_{v}$ , where $b(v)$ denotes the boundary (bulk)  space. A generic quantum state is  thus given as
\begin{equation}
\vert\Psi\rangle=\sum\limits_{b,v} C_{b,v} \vert\chi_{b}\rangle {\otimes} \vert\psi_{v}\rangle ~\label{genstate} 
\end{equation} 
Now, the full Hamiltonian operator ($\widehat{H}$),  operating on $\mathcal{H}$ is given by
\begin{equation}\label{hamil}
\widehat{H}\vert\Psi\rangle=(\widehat{H_{b}}{\otimes}I_{v}+I_{b}{\otimes}\widehat{H_{v}})\vert\Psi\rangle
\end{equation} 
where,  respectively, $I_{b} (I_{v})$ are identity operators on $\mathcal{H}_{b} (\mathcal{H}_{v})$ and $\widehat{H_{b}} (\widehat{H_{v}})$ are the Hamiltonian operators on $\mathcal{H}_{b}(\mathcal{H}_{v})$. 

The first class constraints are realized on Hilbert space as annihilation constraints on physical states. The bulk Hamiltonian operator thus annihilates bulk physical states
\begin{eqnarray}
\widehat{H_{v}}\vert\psi_{v}\rangle=0 \label{bulkham}
\end{eqnarray}

The bulk quantum spacetime is assumed to be free of electric charge and angular momentum, so that eqn. (\ref{bulkham}) is augmented by the relation
\begin{eqnarray}
[\widehat{H_{v}} - \Phi \widehat{Q_{v}} - \Omega \widehat{J_{v}}] | \psi_{v} \rangle = 0 ~.~\label{fullham}
\end{eqnarray}
This assumption gleans from the idea that the generic quantum bulk Hilbert space is invariant under local $U(1)$ gauge transformations and local spacetime rotations (the latter, as part of local Lorentz invariance).

\subsection{Grand Canonical Partition Function}

 We now consider, heuristically, a grand canonical ensemble of quantum spacetimes with horizons as boundaries, in contact with a heat bath, at some (inverse) temperature $\beta$. Strictly speaking, a radiant black hole spacetime must have a {\it Trapping or Dynamical} horizon \cite{ak04} as its inner boundary. However, in our heuristic approach to thermal stability of equilibrium isolated horizons, we overlook the distinction very close to equilibrium. In other words, we ignore the subtleties associated with backreaction of Hawking radiation on quantum bulk geometry. With these assumptions and caveats, the grand canonical partition function is then given as,
\begin{eqnarray}
Z_{G}=Tr(exp(-\beta\widehat{H}+\beta\Phi\widehat{Q}+\beta\Omega\widehat{J})) ~\label{gcpf}
\end{eqnarray}
where  the trace is taken over all states.  This definition,  together with eqn.s (\ref{genstate}) and (\ref{fullham}), yields
\begin{eqnarray}
Z_{G} &=& \sum_{b,v} \vert C_{b,v} \vert^{2} \langle\psi_{v}\vert\psi_{v}\rangle \langle\chi_{b}\vert  exp(-\beta\widehat{H}+\beta\Phi\widehat{Q}+\beta\Omega\widehat{J})  \vert\chi_{b}\rangle \nonumber \\ 
&=& \sum\limits_{b} \vert C_{b} \vert^{2} \langle\chi_{b}\vert exp(-\beta\widehat{H}+\beta\Phi\widehat{Q}+\beta\Omega\widehat{J})  \vert\chi_{b}\rangle ~,~ \label{pf+ham}
\end{eqnarray}
assuming that the bulk states are normalized. The partition function  thus turns out to be  completely determined by the boundary states ($Z_{Gb}$), i.e.,
\begin{eqnarray}
Z=Z_{Gb} &=& Tr_{b} \exp(-\beta\widehat{H}+\beta\Phi\widehat{Q}+\beta\Omega\widehat{J}) \nonumber \\
&=& \sum\limits_{k,l,m} g(k,l,m) \hspace{.1 in} \exp(-\beta(E(A_{k},Q_{l},J_{m})-\Phi Q_{l}-\Omega J_{m})) ~,~\label{bdypf}
\end{eqnarray}
where $g(k,l,m)$ is the degeneracy corresponding to energy $E(A_{k},Q_{l},J_{m})$ and $k,l,m$ are the quantum numbers corresponding to eigenvalues of area, charge and angular momentum respectively.  These quantum numbers are all taken to be discrete \cite{mm14}. Here,  the spectrum of  the boundary Hamiltonian operator is assumed to be  a function of area, charge and angular momentum of the boundary, considered here to be the horizon. Following \cite{ast01}-\cite{mm14}, it is further assumed that these `quantum hairs' all have a discrete spectrum, in parallel with the LQG results. We cannot prove these assumptions at this point, but explore their consequences here. In the Macroscopic area limit $(A_h >> l_P^2 )$ of quantum isolated horizons , they all have 
large eigenvalues i.e. ($k,l,m>>1$), so that, application of the Poisson resummation formula \cite{cm04} gives
\begin{equation}
Z_{G}=\int dx\hspace{.05 in} dy\hspace{.05 in} dz\hspace{.05 in}g(A(x),Q(y),J(z))\hspace{.05 in}  \exp(-\beta(E(A(x),Q(y),J(z))-\Phi Q(y)-\Omega J(z)))
\end{equation}
where $x,y,z$ are respectively the continuum limit of $k,l,m$ respectively. 

We now assume that the macroscopic spectra of the area, charge and angular momentum are {\it linear} in their arguments, so that a change of variables gives, with constant Jacobian, the result
\begin{equation}
Z_{G}=\int dA\hspace{.05 in} dQ\hspace{.05 in} dJ\hspace{.05 in} \exp [S(A)-\beta(E(A,Q,J)-\Phi Q-\Omega J)]~,~ \label{pfresult}
\end{equation}
where, following \cite{ll}, the {\it microcanonical} entropy of the horizon is defined by $\exp S(A) \equiv \frac{ g(A(x),Q(y),J(z))}{\frac{dA}{dx}\frac{dQ}{dy}\frac{dJ}{dz}}$.

\section{Stability Against gaussian Fluctuations}

\subsection{Saddle Point Approximation}

 The equilibrium configuration of black hole is given by the saddle point $\bar{A},\bar{Q},\bar{J}$ in the three dimensional space of integration over area, charge and angular momentum. The idea now is to examine the grand canonical partition function for fluctuations $a=(A-\bar{A}), q=(Q-\bar{Q}),j=(J-\bar{J})$ around the saddle point, in order to determine the stability of the equilibrium isolated horizon under Hawking radiation. We restrict our attention to Gaussian fluctuations, as per common practice in equilibrium statistical mechanics, with the motivation towards extremizing the free energy for the most probable configuration. Taylor expanding eqn (\ref{pfresult}) about the saddle point, yields 
\begin{eqnarray}
Z_{G} &=& \exp[ S(\bar{A})-\beta M(\bar{A},\bar{Q},\bar{J})+\beta\Phi \bar{Q}+\beta\Omega\bar{J}] \nonumber \\
&\times & \int da~ dq~ dj~ \exp \{-\frac{\beta}{2}[( M_{AA}-\frac{S_{AA}}{\beta} )a^{2} + ( M_{QQ})q^{2}+(2 M_{AQ})aq \nonumber \\
&+& ( M_{JJ})j^{2}+(2M_{AJ})aj+(2 M_{QJ})qj] \} ~,~ \label{sadpt}
\end{eqnarray}
where  $M(\bar{A},\bar{Q},\bar{J})$ is the mass of equilibrium isolated horizon. Here $ M_{AQ} \equiv \frac{\partial^2 M}{\partial A \partial Q} \vert_{(\bar{A},\bar{Q},\bar{J})}$ etc. 

We assume just like in Loop Quantum Gravity, observables used here are self-adjoint operators over the boundary Hilbert space, and hence their eigenvalues are real \cite{rov}. It suffices therefore to restrict integrations over the spectra of these operators to the real axes.

Now,  in the Saddle point approximation the coefficients of  terms linear in $a~,~q~,~j$ vanish by definition of the saddle point. These imply that, at saddle point
\begin{equation}\label{beta}
\beta=\frac{S_{A}}{M_{A}}~ ,~ M_{Q}=\Phi ,M_{J}=\Omega
\end{equation}

\subsection{Stability Criteria}

Convergence of the integral (\ref{sadpt}) implies  that the Hessian matrix ($H$) has to be positive definite, where
\begin{eqnarray}
 H = \left( \begin{array}{ccc}
\beta M_{AA}(\bar{A},\bar{Q},\bar{J})- S_{AA}(\bar{A}) \hspace{.3 in}& \beta M_{AQ}(\bar{A},\bar{Q},\bar{J}) \hspace{.3 in} & \beta M_{AJ}(\bar{A},\bar{Q},\bar{J})\vspace{.1 in} \\  
\beta M_{AQ}(\bar{A},\bar{Q},\bar{J}) \hspace{.3 in}& \beta M_{QQ}(\bar{A},\bar{Q},\bar{J}) \hspace{.3 in}& \beta M_{JQ}(\bar{A},\bar{Q},\bar{J}) \vspace{.1 in} \\
\beta M_{AJ}(\bar{A},\bar{Q},\bar{J}) \hspace{.3 in}& \beta M_{JQ}(\bar{A},\bar{Q},\bar{J}) \hspace{.3 in}& \beta M_{JJ}(\bar{A},\bar{Q},\bar{J}) \end{array} \right) \label{hess}
\end{eqnarray}
The necessary and sufficient conditions for  a real symmetric square matrix to be positive definite are : 'determinants all principal square submatrices, and the determinant of the full matrix, are positive.'\cite{meyer} This condition leads to the following `stability criteria' :
\begin{eqnarray}
 M_{AA}(\bar{A},\bar{Q},\bar{J})- \frac{S_{AA}(\bar{A})}{\beta} & > & 0 \label{stab1} \\ 
\vspace{.1 in}  M_{QQ}(\bar{A},\bar{Q},\bar{J}) &>& 0 \label{stab2}  \\
 M_{JJ}(\bar{A},\bar{Q},\bar{J}) &>& 0 \label{stab3}  \\ 
M_{QQ}(\bar{A},\bar{Q},\bar{J})M_{JJ}(\bar{A},\bar{Q},\bar{J}) & - & (M_{JQ}(\bar{A},\bar{Q},\bar{J}))^{2}>0 \label{stab4} \\
M_{JJ}(\bar{A},\bar{Q},\bar{J})\big( M_{AA}\big(\bar{A},\bar{Q},\bar{J})- \frac{S_{AA}(\bar{A})}{\beta}\big) &-& (M_{AJ}(\bar{A},\bar{Q},\bar{J}))^{2}>0 \label{stab5} \\
M_{QQ}(\bar{A},\bar{Q},\bar{J})\big( M_{AA}\big(\bar{A},\bar{Q},\bar{J})- \frac{S_{AA}(\bar{A})}{\beta}\big) &-&  (M_{AQ}(\bar{A},\bar{Q},\bar{J}))^{2}>0 \label{stab6} 
\end{eqnarray}

$[\big( M_{AA}(\bar{A},\bar{Q},\bar{J})- \frac{S_{AA}(\bar{A})}{\beta}\big)(M_{QQ}(\bar{A},\bar{Q},\bar{J})M_{JJ}(\bar{A},\bar{Q},\bar{J}) -(M_{JQ}(\bar{A},\bar{Q},\bar{J}))^{2})$\\

$- M_{AQ}(\bar{A},\bar{Q},\bar{J})(M_{AQ}(\bar{A},\bar{Q},\bar{J})M_{JJ}(\bar{A},\bar{Q},\bar{J})-M_{JQ}(\bar{A},\bar{Q},\bar{J})M_{AJ}(\bar{A},\bar{Q},\bar{J}))$\
\begin{equation}
+ M_{AJ}(\bar{A},\bar{Q},\bar{J})(M_{AQ}(\bar{A},\bar{Q},\bar{J})M_{JQ}(\bar{A},\bar{Q},\bar{J})-M_{QQ}(\bar{A},\bar{Q},\bar{J})M_{AJ}(\bar{A},\bar{Q},\bar{J}))] > 0 \label{stab7}
\end{equation}
Of course, (inverse) temperature $\beta$ is assumed to be positive for a stable configuration.\\  

Now, the temperature is defined as $T \equiv \frac{1}{\beta}$; eqn. (\ref{beta}) implies that $T= \frac{M_{A}}{S_{A}}$. Eqn.s (\ref{kment}) and (\ref{bek}) together yield $S_{A}= \frac{1}{4A_{P}} - \frac{3}{2A}$ and is positive for macroscopic black holes as $A >> A_{P}$. So, positivity of $M_{A}$ implies the positivity of $\beta$ for macroscopic black holes. The relation $T= \frac{M_{A}}{S_{A}}$ implies $\frac{dT}{dA}= \frac{M_{A}}{(S_{A})^{2}}(\beta M_{AA}- S_{AA})$.  So, what is new is the requirement that this temperature must increase with horizon area, inherent in the positivity of the quantity ($\beta M_{AA} - S_{AA}$) which appears in several of the stability criteria. If this is violated, as for example in case of the standard Schwarzschild black hole \cite{cm04}, thermal instability is inevitable.

The convexity property of the entropy follows from the condition of convergence of partition function under gaussian fluctuations \cite{cm04}, \cite{ll}, \cite{mon}.  The thermal stability is related to the convexity property of entropy. Hence, the above conditions are correctly the conditions for thermal stability. For chargeless, non-rotating horizons, eqn. (\ref{stab1}) reproduces the thermal stability criterion and condition of positive specific heat(i.e.variation of black hole mass with temperature) given in \cite{pm07}, as expected. Actually for a chargeless, non-rotating black hole, both the mass and the temperature are functions of the horizon area ($A$) only. From these one can define the specific heat as $C = \frac{dM}{dT}= \frac{(S_{A})^{2}}{(\beta M_{AA}- S_{AA})}$.

For charged, non-rotating black holes, eqn.s (\ref{stab1}), (\ref{stab2}) and (\ref{stab6}) describe the stability, in perfect agreement with \cite{mm12}, while (\ref{stab1}), (\ref{stab3}) and (\ref{stab5}) describe the thermal stability criteria for uncharged rotating radiant horizons. The new feature for black holes with both charge and angular momentum is that not only does the specific heat has to be positive for stability, but the charge and the angular momentum play important roles as well. 

As claimed in the Introduction, the thermal stability criteria above are derived by the application of standard statistical mechanical formalism to a quantum horizon characterized by various observables having discrete eigenvalue spectra. Thus, no aspect of classical geometry enters the derivation of these criteria. Given the classical metrics specifying various classical black hole spacetimes, the mass can be obtained as an explicit function of the area, charge and angular momentum of the horizon. It is then possible, on the basis of our stability criteria, to {\it predict} which classical black holes will radiate away to extinction, and which ones might find some stability, and for what range of  parameters. This is what is attempted in the next section.

{\section{Predicting Thermal Stability of Classical Black Holes}

 Notice that in the stability criteria derived in the last section, first and second order derivatives of the microcanonical entropy of the horizon at equilibrium play a crucial role, in making some of the criteria non-trivial. Thus, corrections to the microcanonical entropy beyond the Bekenstein-Hawking area law, arising due to quantum spacetime fluctuations might play a role of some significance. It has been shown that \cite{km00} the microcanonical entropy for  {\it macroscopic} isolated horizons has the form
\begin{eqnarray}
S~&=&~S_{BH} ~-~\frac32 \log S_{BH} +{\cal O}(S_{BH}^{-1}) ~\label{kment} \\
S_{BH} ~&=& ~ \frac{A_h}{4 A_P}~,~A_P \equiv {\rm Planck~area} ~. \label{bek}
\end{eqnarray}

\subsection{Kerr-Newman Black Hole}

The Kerr-Newman metric of asymptotically flat Black Hole is given in Boyer-Lindquist coordinates as
\begin{equation}\label{knmetric}
ds^{2}= -\frac{\Xi}{\Sigma}(dt-a\hspace{.02 in} sin^{2}\theta\hspace{.02 in} d\phi)^{2} +\frac{sin^{2}\theta}{\Sigma}((r^{2}+a^{2}) d\phi -a\vspace{.02 in}dt)^{2} +\frac{\Sigma}{\Xi}dr^{2}+ \Sigma d\theta ^{2}
\end{equation}
where, $ \Xi= r^{2}-2\hspace{.02 in}M \hspace{.02 in}r + a^{2}+Q^{2} ,\hspace{.2in} \Sigma= r^{2}+a^{2}\hspace{.02 in}cos^{2}\theta  , \hspace{.2 in}  a=\frac{J}{M} $.
The generalized Smarr formula for the  Kerr-Newman Black Hole is given as \cite{cck} 
\begin{equation}\label{knmass}
M^{2}=\frac{A}{16\pi}+\frac{\pi}{A}(4J^{2}+Q^{4})+\frac{Q^{2}}{2}
\end{equation}\\

Calculation, done in the subsection (\ref{appenkn}), shows that there does not exist any single set of value $(A,J,Q)$ such that the conditions (\ref{stab1} - \ref{stab7}) hold simultaneously. So, this kind of black hole is thermally unstable under hawking radiation.

\subsection{AdS Black Holes}

The thermal instability discerned in the last subsection for the standard asymptotically flat general relativistic black hole spacetimes raises the question as to whether the asymptotically Anti-de Sitter versions of these spacetimes are thermally more stable, for some region of their parameter space, as has been noticed decades ago by Hawking and Page \cite{hawp} within a semiclassical approach. In contrast, the quantum geometry underpinning of our analysis in the previous sections is independent of specific black hole metrics, giving us very general criteria for thermal stability. The mass-horizon area functional dependence derived from classical metrics of specific black holes therefore permits predictions of stability behaviour of specific black holes under Hawking radiation. In this sense, classical geometry provides us with fiducials for verification of validity of the stability criteria derived earlier. This motivates their application to AdS black holes in this subsection. 

The AdS Kerr-Newman black hole is given in Boyer–Lindquist coordinates as 
\begin{equation}\label{adsknmetric}
ds^{2}= -\frac{\Delta_{r}}{\rho^{2}}(dt-\frac{a\hspace{.02 in} sin^{2}\theta}{\Sigma}\hspace{.02 in} d\phi)^{2} +\frac{\Delta_{\theta}\hspace{.02 in} sin^{2}\theta}{\rho^{2}}(\frac{r^{2}+a^{2}}{\Sigma} d\phi -a\vspace{.02 in}dt)^{2} +\frac{\rho^{2}}{\Delta_{r}}dr^{2}+ \frac{\rho^{2}}{\Delta_{\theta}} d\theta ^{2}
\end{equation}
where, $ \Sigma= 1-\frac{a^{2}}{l^{2}},\hspace{.1 in} \Delta_{r}=(r^{2}+ a^{2})(1+\frac{r^{2}}{l^{2}})-2\hspace{.02 in}M \hspace{.02 in}r +Q^{2} ,\hspace{.1 in}\Delta_{\theta}= 1-\frac{a^{2}cos^{2}\theta}{l^{2}},\hspace{.1in} \rho^{2}= r^{2}+a^{2}\hspace{.02 in}cos^{2}\theta  , \hspace{.1 in}  a=\frac{J}{M} $. The generalized Smarr formula for the AdS Kerr-Newman Black Hole is given as \cite{cck} 
\begin{equation}\label{adsknmass}
M^{2}=\frac{A}{16\pi}+\frac{\pi}{A}(4J^{2}+Q^{4})+\frac{Q^{2}}{2}+\frac{J^{2}}{l^{2}}+\frac{A}{8\pi l^{2}}(Q^{2}+\frac{A}{4\pi}+\frac{A^{2}}{32\pi^{2}l^{2}})
\end{equation}
where  the cosmological constant ($\Lambda$) is defined in terms of a cosmic length parameter as $\Lambda = -1/l^2$.\\

Calculation, done in the subsection(\ref{appeadskn}), shows that conditions (\ref{stab1} - \ref{stab7}) holds simultaneously only if $\frac{A}{l^{2}}$ is greater than $\frac{J}{A}, \frac{Q^{2}}{A}$ , infact greater by one order of magnitude. So, this kind of black hole is thermally stable under hawking radiation only within the range as stated.

\subsection{Asymptotically Flat String Theoretic Black Hole}

Here we consider the low energy effective field theory describing heterotic string theory, which describes a black hole carrying finite
amount of charge and angular momentum \cite{sen}. In low energy limit the effective four dimensional theory contains gravity, maxwell field, dilaton field and antisymmetric gauge field.The solution of the metric turns out be a black hole whose charge, mass and angular momentum are determined by various fundamental parameters of the theory \cite{sen}. The classical metric for such a black hole is 
\begin{eqnarray}
ds^2 &=& -\frac{\rho^2 +a^2\cos^2\theta -2m\rho}{ \rho^2+a^2\cos^2\theta +2m\rho\sinh^2{\alpha\over 2}} dt^2 +{\rho^2 +a^2\cos^2\theta +2m\rho\sinh^2{\alpha\over 2}\over \rho^2 +a^2 -2m\rho}d\rho^2 \nonumber \\
&+& (\rho^2+a^2\cos^2\theta +2m\rho\sinh^2{\alpha\over 2}) d\theta^2 -{4m\rho a\cosh^2{\alpha\over 2}\sin^2\theta\over \rho^2
+a^2\cos^2\theta +2m\rho\sinh^2{\alpha\over 2}} dtd\phi \nonumber \\
&+& \{(\rho^2+a^2)(\rho^2+a^2\cos^2\theta) +2m\rho a^2\sin^2\theta +4m\rho
(\rho^2+a^2) \sinh^2{\alpha\over 2}+ 4m^2\rho^2\sinh^4{\alpha\over 2}\} \nonumber \\
& \times &  {\sin^2\theta \over \rho^2+a^2\cos^2\theta +2m\rho\sinh^2{\alpha\over 2}} d\phi^2 \label{senmetric}
\end{eqnarray}

This metric describes a black hole solution with mass $M$, charge $Q$, and angular momentum $J$ given by 
\begin{equation}\label{senparameter}
M={m\over 2} (1+\cosh\alpha), \hspace{.2 in}Q={m\over\sqrt 2}\sinh\alpha, \hspace{.2 in} J={ma\over 2} (1+\cosh\alpha)
\end{equation}
It will be more convenient to express $m$, $a$ and
$\alpha$ in terms of the independent physical parameters $M$, $J$ and $Q$
by inverting the relations given in( \ref{senparameter}). We get
\begin{eqnarray}
m=M-{Q^2\over 2M}, \hspace{.2 in}\sinh\alpha ={2\sqrt 2 QM\over 2M^2 -Q^2},\hspace{.2 in} a={J\over M} \label{malpha}
\end{eqnarray}
The area of the horizon turns out to be 
\begin{eqnarray}
A=8\pi M\Bigg(M-\frac{Q^{2}}{2M}+\sqrt{(M-\frac{Q^{2}}{2M})^{2}-\frac{J^{2}}{M^{2}}} \Bigg) \label{senarea}
\end{eqnarray}
which gives the mass of the black hole($M$) as
\begin{eqnarray}
M^{2}=\frac{A}{16\pi}+\frac{Q^{2}}{2}+\frac{4\pi J^{2}}{A} \label{senmass}
\end{eqnarray}

Calculation, done in the subsection (\ref{appeks}), shows that there does not exist any single set of value $(A,J,Q)$ such that the conditions (\ref{stab1} - \ref{stab7}) hold simultaneously. So, this kind of black hole is thermally unstable under hawking radiation.

\subsection{ Five Dimensional Asymptotically Flat Dilatonic Black Hole With Rotation}

Here we consider the dilaton field coupled to gravity in presence of the Maxwell field in five dimension \cite{hor}. Such solutions are derived from the standard four dimensional Kerr solution of Einstein's equation, by constructing the five dimensional product space obtained by tensoring the Kerr spacetime with  $R$. Boosting the Kerr solution along the real line thus gives a rotating charged black hole in a five dimensional Lorentzian spacetime. So, this kind of black hole has $\big((S^{2} \times R^{2}) \times R \big) $ structure , while kerr-newman black hole has $(S^{2} \times R^{2})$ structure. We can intuitively conclude , from the above sturctural similarity , that Five Dimensional Asymptotically Flat Dilatonic Black Hole With Rotation are unstable under hawking radiation like kerr-newman black hole. Infact detail calculation, done in the subsection (\ref{appefd}), meets our intuition as well.
 
\section{Summary and Discussion}

In the following table, we summarize the work of previous sections, for a clearer perspective.
\vglue .3cm
\begin{tabular}{|l|l|}
\hline
{\bf Type Of Black Hole} & {\bf Whether Stable}  \\
\hline
Asymptotically Flat Black Holes & Unstable  \\
\hline
ADS Black Holes & Stable if $\frac{A}{l^{2}}$ is greater than $\frac{J}{A} , \frac{Q^{2}}{A}$ by one order of magnitude \\
\hline
String Theoretic Black Holes & Unstable \\
\hline
Dilaton Black Holes without Cosmological Constant & Unstable \\
\hline
\end{tabular}
\vglue .3cm

We reiterate that our analysis is quite independent of specific classical spacetime geometries, relying as it does on quantum aspects of spacetime. The construction of the partition function used standard formulations of equilibrium statistical mechanics augmented by results from canonical Quantum Gravity, with extra inputs regarding the behaviour of the microcanonical entropy as a function of area {\it beyond the Bekenstein-Hawking area law}, as for instance derived from Loop Quantum Gravity \cite{km00}. However, we emphasize that the results are more general than being restricted to any specific proposal for quantum spacetime geometry, requiring only certain functional dependences on horizon area and other parameters of statistical mechanical quantities like entropy. It also stands to reason that our stability criteria are useful for predicting the thermal behaviour vis-a-vis Hawking radiation for specific astrophysical black holes. In particular, our criteria precisely predict regions of the parameter space of specific black hole solutions, not only in general relativity, but also of extensions inspired from warped geometries and string theories, where these solutions are stable under Hawking radiation.

It is also noteworthy that the approach is useful for making predictions on the thermal stability of black holes in Lorentzian spacetimes with arbitrary number of spatial dimensions. It can also be generalized to black holes with arbitrary `hairs' (charges) - either quantum or classical \cite{sinha}. 

There are however, subtleties of a statistical mechanical nature which have not been addressed in this paper. The most important of these is the {\it nature} of the thermal instability discerned by us. While there are indications that the instability in most cases can be associated with some sort of phase transition \cite{pm07}, the very general approach here has not yet been applied to discuss the full range of thermal behaviour exhibited specifically for AdS Schwarzschild black holes, for instance, as discussed in detail in \cite{hawp}. Crucially, there are `phases' discussed in that paper which have not been fully explored via our more `quantum geometry' approach, as distinct from the semiclassical approach employed in \cite{hawp}. We hope to return to these important issues in a future publication.

\vspace{.2 in}

\section{Appendix}

In this section, we will show the detail calculation for Kerr-Newman Black Hole , AdS Black Holes , Asymptotically Flat String Theoretic Black Hole and  Five Dimensional Asymptotically Flat Dilatonic Black Hole With Rotation seperately.\\

For the first three kinds of black holes, as stated above, mass($M$) is expressed as $M^{2}= M^{2}(A,Q,J)$ with $M^{2}|_{JQ}= 0$ , where $M^{2}|_{JQ} \equiv \frac{\partial^{2}M^{2}}{\partial J \partial Q}$. It is also a common feature of the expression of $M^{2}$ for all these three types of black holes that $M^{2}$ is analytic in $A,Q$ and $J$. So, it is better to express various derivatives that appears in (\ref{stab1} - \ref{stab7}) in terms of derivatives of $M^{2}$ .\\

$M^{2}= M^{2}(A,Q,J)$ gives $M_{A} = \frac{M^{2}|_{A}}{2M}$ , $M_{AQ} = \frac{1}{4M^{3}}{ \big(2M^{2}M^{2}|_{AQ} - (M^{2}|_{A} M^{2}|_{Q})\big)}$ etc. \\

\subsection{Results for Kerr-Newman Black Hole} \label{appenkn}
$ S_{A} = \frac{1}{4A_{P}} - \frac{3}{2A}$ , $S_{AA} = \frac{3}{2A^{2}}$ , $M^{2}|_{A} = \frac{1}{16\pi} - \frac{\pi}{A^{2}}(4J^{2}+Q^{4})$ , $M^{2}|_{AA}= \frac{2\pi}{A^{3}}(4J^{2}+Q^{4})$ , $M^{2}|_{Q}= Q + \frac{4\pi Q^{3}}{A}$ , $M^{2}|_{QQ}= \frac{12\pi Q^{2}}{A}$ , $M^{2}|_{J}= \frac{8\pi J}{A}$ , $M^{2}|_{JJ}= \frac{8\pi}{A}$ , $M^{2}|_{AJ}= -\frac{8\pi J}{A^{2}}$ , $M^{2}|_{AQ}= - \frac{4\pi Q^{3}}{A^{2}}$ , $M^{2}|_{JQ}= 0$\\

We write a small computer program and ask the computer to do the calculation of left hand sides of (\ref{stab1} - \ref{stab7}) on basis of the above input. We scan over various range of all possible range of $(A,Q,J)$ and find not single point such that all the left hand sides of (\ref{stab1} - \ref{stab7}) are positive simultaneously.  The process of calculation is described as follows,\\

$A_{P} \sim 10^{-70} m^{2}$ and for a typical black hole, $A\sim 10^{5} m^{2}$. In fact this value is even larger for macroscopic black holes. $\therefore \frac{A}{A_{P}} \sim 10^{75}$ and this implies $S_{A} = \frac{1}{4A_{P}} - \frac{3}{2A}$ is positive for macroscopic black holes. So, positivity of temperature ($\frac{1}{\beta} = \frac{M_{A}}{S_{A}}$) implies the positivity of $M_{A}$ i.e. positivity of $M^{2}|_{A} (= 2MM_{A})$. So, 
$M^{2}|_{A} = \frac{1}{16\pi} - \frac{\pi}{A^{2}}(4J^{2}+Q^{4}) > 0$ implies $\frac{J}{A}< \frac{1}{8\pi}$ and $\frac{Q^{2}}{A}< \frac{1}{4\pi}$. \\

Now, the expression (\ref{knmass}) gives $[M^{2}]=[A]=[Q^{2}]=[J]=[A_{P}]$, where $[A]$ is the dimension of $A$ etc.  \hspace{1 in} Define, $\frac{J}{A} \equiv x , \frac{Q^{2}}{A} \equiv y , \frac{A}{A_{P}} \equiv z$ and these $x,y,z$ are dimensionless. So, we have $M^{2}|_{A}= \frac{1}{16\pi} - 4\pi x^{2} -\pi y^{2}$ , $AM^{2}|_{A}= 8\pi x^{2} +2\pi y^{2} $, $M^{2}|_{J}= 8\pi x$ , $AM^{2}|_{JJ}= 8\pi$ , $\frac{M^{2}|_{Q}}{\sqrt{A}}= y^{1/2} + 4\pi y^{3/2}$ , $M^{2}|_{QQ}= 1 + 12\pi y$ , $AM^{2}|_{AJ}= -8\pi x$ , $\sqrt{A}M^{2}|_{AQ} = -4\pi y^{3/2}$ \\

Now, $\beta$ and left hand sides of (\ref{stab1} - \ref{stab7}) can be described interms of the above dimensionless functions as,\\

$\beta$= $\frac{f_{1}}{2M}$ , where $f_{1}= M^{2}|_{A}$ \\

left hand side of \ref{stab1}= $\frac{1}{4M^{3}} \cdot f_{2}$, where $f_{2}= \big(2 \cdot\frac{M^{2}}{A} \cdot AM^{2}|_{AA} - (M^{2}|_{A})^{2}\big) - 12M^{2}|_{A} \cdot \frac{M^{2}}{A} \cdot \frac{1}{z-6} $ \\

left hand side of \ref{stab2}= $\frac{ A}{4M^{3}} \cdot f_{3}$, where $f_{3}= \big(2 \cdot\frac{M^{2}}{A} \cdot M^{2}|_{QQ} - (\frac{M^{2}|_{A}}{\sqrt A})^{2}\big) $ \\

left hand side of \ref{stab3}= $\frac{1}{4M^{3}} \cdot f_{4}$, where $f_{4}= \big(2 \cdot\frac{M^{2}}{A} \cdot A M^{2}|_{JJ} - (M^{2}|_{J})^{2}\big) $ \\

left hand side of \ref{stab4}= $\frac{A }{16M^{6}} \cdot \big (f_{3} \cdot f_{4} - (f_{5})^{2}\big)$, where $f_{5}= \big(2 \cdot\frac{M^{2}}{A} \cdot \sqrt{A} M^{2}|_{JQ} - \frac{M^{2}|_{Q}}{\sqrt A} \cdot M^{2}|_{J} \big) $ \\ 

left hand side of \ref{stab5}= $\frac{1}{16M^{6}} \cdot \big (f_{4} \cdot f_{2} - (f_{6})^{2}\big)$, where $f_{6}= \big(2 \cdot\frac{M^{2}}{A} \cdot {A} M^{2}|_{JA} - M^{2}|_{A} \cdot M^{2}|_{J} \big) $ \\

left hand side of \ref{stab6}= $\frac{A}{16M^{6}} \cdot \big (f_{3} \cdot f_{2} - (f_{7})^{2}\big)$, where $f_{7}= \big(2 \cdot\frac{M^{2}}{A} \cdot \sqrt{A} M^{2}|_{QA} - M^{2}|_{A} \cdot \frac{M^{2}|_{Q}}{\sqrt{A}} \big) $ \\

left hand side of \ref{stab7}= $\frac{1}{16AM^{2}} \cdot \Big(f_{2}\cdot \big(f_{3}f_{4} - (f_{5})^{2}\big) - f_{7} \cdot \big(f_{4}f_{7} - f_{5}f_{6}\big) + f_{6} \cdot \big(f_{5}f_{7} - f_{3}f_{6}\big)\Big)$\\

All these functions $f_{1},...,f_{7}$ are dimensionless functions of $x,y,z$. Now, we run our computer program to find the points($x,y$) within the range $0<x<\frac{1}{8\pi}$ and $0<y<\frac{1}{4\pi}$ such that left hand sides of (\ref{stab1} - \ref{stab7}) are simultaneously positive. But, we do not get a single such point. It is true for any value of $z$ greater than $6$, as $z>6$ implies $S_{A}$ is positive.

\subsection{Results for ADS Kerr-Newman Black Hole} \label{appeadskn}

$ S_{A} = \frac{1}{4A_{P}} - \frac{3}{2A}$ , $S_{AA} = \frac{3}{2A^{2}}$ , $M^{2}|_{A}= \frac{1}{16\pi}-\frac{\pi}{A^{2}}(4J^{2}+Q^{4})+\frac{Q^{2}}{8\pi l^{2}} + \frac{A}{16\pi^{2}l^{2}} + \frac{3A^{2}}{256\pi^{3}l^{4}}$  \hspace{1.5 in} $ M^{2}|_{AA}= \frac{2\pi}{A^{3}}(4J^{2}+Q^{4})+\frac{1}{16\pi^{2}l^{2}} + \frac{3A}{128\pi^{3}l^{4}}$ , $M^{2}|_{Q}= Q+ \frac{4\pi Q^{3}}{A} + \frac{AQ}{4\pi l^{2}}$ , $M^{2}|_{QQ}= 1+ \frac{12\pi Q^{2}}{A} + \frac{A}{4\pi l^{2}}$ \hspace{1 in} $M^{2}|_{J} = \frac{8\pi J}{A} + \frac{2J}{l^{2}}$ , $M^{2}|_{JJ}= \frac{2}{l^{2}} + \frac{8\pi}{A}$ , $M^{2}|_{AJ}= -\frac{8\pi J}{A^{2}}$ , $M^{2}|_{AQ}= \frac{Q}{4\pi l^{2}} - \frac{4\pi Q^{3}}{A^{2}}$ , $M^{2}|_{JQ}= 0$ \\

Define, $\frac{A}{l^{2}} \equiv u, \frac{J}{A} \equiv x , \frac{Q^{2}}{A} \equiv y , \frac{A}{A_{P}} \equiv z$. So, we have $M^{2}|_{A}= \frac{1}{16\pi} - 4\pi x^{2} -\pi y^{2} + \frac{yu}{8\pi} + \frac{u}{16\pi^{2}} + \frac{3u^{2}}{256\pi^{3}}$ , $AM^{2}|_{AA}= 8\pi x^{2} +2\pi y^{2} + \frac{u}{16\pi^{2}} + \frac{3u^{2}}{128\pi^{3}}$, $M^{2}|_{J}= 8\pi x + 2xu$ , $AM^{2}|_{JJ}= 8\pi + 2u$ , $\frac{M^{2}|_{Q}}{\sqrt{A}}= y^{1/2} +  4\pi y^{3/2} + \frac{uy^{1/2}}{4\pi}$ , $M^{2}|_{QQ}= 1 + 12\pi y + \frac{u}{4\pi}$ , $AM^{2}|_{AJ}= -8\pi x$ , $\sqrt{A}M^{2}|_{AJ} = -4\pi y^{3/2} + \frac{uy^{1/2}}{4\pi}$ \\

Like earlier(case of kerr-Newman black hole), we can define $f_{1}, ... ,f_{7}$ interms of $x,y,z,u$. We can also express  $\beta$ and left hand sides of (\ref{stab1} - \ref{stab7}) interms of $f_{1}, ... ,f_{7}$ as before. We then run our computer program to find the points($x,y,u$)  such that left hand sides of (\ref{stab1} - \ref{stab7}), $\beta$ are simultaneously positive. We find (as shown in the table below)such points only if $u> x,y$; by one order of magnitude. It is true for any value of $z$ greater than $6$, as argued earlier. 

\vglue .3cm
\begin{tabular}{|l|l|l|}
\hline
{\bf Value of $ u(=\frac{A}{l^{2}})$}&{\bf Value of $\frac{x}{u}(=\frac{(J/A)}{(A/l^{2})})$}&{\bf Value of $\frac{y}{u}(=\frac{(Q^{2}/A)}{(A/l^{2})})$} \\
\hline
$1$ & $9.99 \times 10^{-3}$ & $8.99 \times 10^{-2}$ \\
\hline
$10^{1}$ & $3.00 \times 10^{-3}$ & $2.60 \times 10^{-2}$ \\
\hline
$10^{2}$ & $8.99 \times 10^{-4} $ & $1.96 \times 10^{-2}$ \\
\hline
$10^{3}$ & $8.19\times 10^{-4}$ & $1.89 \times 10^{-2}$ \\
\hline
$10^{4}$ & $7.75 \times 10^{-4}$ & $ 1.88 \times 10^{-2}$\\
\hline
$10^{5}$ & $7.24 \times 10^{-4}$ & $1.87 \times 10^{-2}$ \\
\hline
\end{tabular}
\vglue .3cm

This table shows the selected six points in the ($u, \frac{x}{u}, \frac{y}{u}$) space ,  such that ADS KN black hole is stable in these points. This table ofcourse shows the maximum possible values of $\frac{x}{u}, \frac{y}{u}$ for a given value of '$u$' within the region of stability.\\

The generalized Smarr formula for the AdS Kerr-Newman Black Hole \ref{adsknmass} takes  the form of the generalized Smarr formula for the  Kerr-Newman Black Hole \ref{knmass}  in the limit $l \rightarrow \infty$. These two equations start to differ gradually as $l$ starts to become smaller. Kerr-Newman black holes are thermally unstable. Hence ADS Kerr-Newman black hole can have stability, but not for large values of '$l$'. This fact is relected in the table above. We scan the entire space of $(u,x,y)$ and find that increment of $ u(=\frac{A}{l^{2}})$ decreses the values of $ x(=\frac{J}{A})$ and $ y(=\frac{Q^{2}}{A})$ to attain thermal stability of ADS KN black hole.

\subsection{Results for Asymptotically Flat String Theoretic Black Hole} \label{appeks}

$ S_{A} = \frac{1}{4A_{P}} - \frac{3}{2A}$ , $S_{AA} = \frac{3}{2A^{2}}$ , $M^{2}|_{A} = \frac{1}{16\pi} - \frac{4\pi J^{2}}{A^{2}}$ , $M^{2}|_{AA}= \frac{8\pi J^{2}}{A^{3}}$ , $M^{2}|_{Q}= Q $ , $M^{2}|_{QQ}= 1$ , $M^{2}|_{J}= \frac{8\pi J}{A}$ , $M^{2}|_{JJ}= \frac{8\pi}{A}$ , $M^{2}|_{AJ}= -\frac{8\pi J}{A^{2}}$ , $M^{2}|_{AQ}= 0$ , $M^{2}|_{JQ}= 0$\\

So, we have $M^{2}|_{A}= \frac{1}{16\pi} - 4\pi x^{2} $ , $AM^{2}|_{AA}= 8\pi x^{2}$, $M^{2}|_{J}= 8\pi x$ , $AM^{2}|_{JJ}= 8\pi$ , $\frac{M^{2}|_{Q}}{\sqrt{A}}= y^{1/2} $ , $M^{2}|_{QQ}= 1 $ , $AM^{2}|_{AJ}= -8\pi x$ , $\sqrt{A}M^{2}|_{AQ} = 0$ \\

Here, $M^{2}|_{A} = \frac{1}{16\pi} - \frac{4\pi J^{2}}{A^{2}} > 0$ implies $\frac{J}{A}< \frac{1}{8\pi}$ and is the condition for positivity of $\beta$. Now, we define $f_{1}, ... ,f_{7}$ as before interms of $x,y,z$ and express left hand sides of (\ref{stab1} - \ref{stab7}) interms of $f_{1}, ... ,f_{7}$ as before. We then run our computer program to find the points($x,y$)  such that left hand sides of (\ref{stab1} - \ref{stab7}) are simultaneously positive within the range $0<x<\frac{1}{8\pi}$.  But, we do not get a single such point. It is true for any value of $z$ greater than $6$, as argued earlier.

\subsection{Results for  Five Dimensional Asymptotically Flat Dilatonic Black Hole With Rotation} \label{appefd}

Let, $m~,~a$ be the mass and rotation parameter of the original Kerr solution and $v$ is the velocity of the boost in the extra direction.

The resulting metric is as,
\begin{equation}\label{d1metric}
ds^2 =  - \frac{1-Z}{B} dt^2- \frac{2 aZ \sin^2 \theta}{ B \sqrt{1 - v^2}}dt d\phi
  + \left[ B (r^2 + a^2) + a^2\sin^2\theta {Z \over B} \right]\sin^2 \theta d\phi^2
     + B { \Sigma \over \Delta_0} dr^2 + B \Sigma  d\theta^2 
\end{equation}     
where
$ B = \sqrt{ 1 + { v^2 Z \over 1 - v^2}}~,~ \hspace{.2 in} Z = { 2 m r \over \Sigma},\hspace{.2 in}\Delta_0 = r^2 + a^2 - 2 m r ,\hspace{.2 in}\Sigma = r^2 + a^2 \cos^2 \theta $. The dilaton field is given by $\phi = (-\sqrt{3}/2) \log B$. 

This gives the mass($M$), charge($Q$) and angular momentum ($J$)of the black hole as,
\begin{eqnarray}
M &=& m \left( 1+\frac{v^{2}}{2(1-v^{2})} \right) \label{d1mass} \\
Q &=& \frac{mv}{1-v^{2}} \label{d1charge} \\
J &=& \frac{ma}{\sqrt{1-v^{2}}} \label{d1angumome}
\end{eqnarray}
One can solve eqn.s (\ref{d1mass}) and (\ref{d1charge}) for the boost velocity $v$ in terms of these parameters
\begin{equation}\label{d1velocity}
v=\sqrt{2+(\frac{M}{Q})^{2}}-\frac{M}{Q}
\end{equation}
Similarly, eqn.s (\ref{d1mass}), (\ref{d1charge}) and (\ref{d1velocity}) can be inverted to yield $m$ as a function of these parameters
\begin{eqnarray} m = \frac{3M}{2}-\frac{\sqrt{M^{2}+2Q^{2}}}{2} ~.~\label{dim}
\end{eqnarray}
The area of the black hole $A$ is then given as \cite{hor}
\begin{equation}\label{d1area}
A=8\pi \left[ C+\sqrt{C^{2}-J^{2}} \right]
\end{equation}
where, 
\begin{eqnarray}
C &=& \frac{m^{2}}{\sqrt{1-v^{2}}} \nonumber \\
 &=& \frac{\frac{9M^{2}Q}{4}-\frac{3M^{2}Q}{2}\Big(\sqrt{1+\frac{2Q^{2}}{M^{2}}}\Big)+\frac{M^{2}Q+2Q^{3}}{4}}{\sqrt{2M^{2}\Big(\sqrt{1+\frac{2Q^{2}}{M^{2}}}\Big)-2M^{2}-Q^{2}}} \label{dlc}
\end{eqnarray}

While eqn. (\ref{d1mass}) expresses the mass of the black hole as a function of the parameters $v$ and $m$, eliminating the latter in terms of the parameters $A, Q, J$ is a complicated algebraic task, involving the inversion of eqn. (\ref{dlc}). An easier approach is to rederive the stability criteria from the Grand Canonical Partition Function and evaluate the saddle point integrals over the variables $m, v, a$, taking the appropriate Jacobian into account.

Assume,    
\begin{eqnarray}\label{cov}
\cosh \eta = \frac{1}{\sqrt{1-v^{2}}} \hspace{.05 in}, \hspace{.1 in} \cos \mu =\frac{a}{m} 
\end{eqnarray}
Eqn.s (\ref{d1mass}), (\ref{d1charge})(\ref{d1angumome}) (\ref{d1area})and (\ref{cov}) together imply
\begin{eqnarray}\label{repara}
M=m\Big(1+\frac{\sinh^{2} \eta}{2}\Big) &,& J=m^{2} \cos \mu \cosh \eta \nonumber \\
Q=\frac{m}{2} \sinh 2\eta &,& A=8\pi m^{2} \cosh \eta \Big(1+\sin \mu \Big) \label{param}
\end{eqnarray}
Eqn. (\ref{repara}) implies that large area ($A$) means large value of '$m$' and '$\cosh \eta$'.\\

Now, we will calculate the Grand Canonical Partition function($Z_{G}$) interms of the new variables $(m,\eta,\mu)$. \

Let us assume that the saddle point be at $(\Bar{m},\Bar{\eta},\Bar{\mu})$ and define the fluctuations around this point as
 \begin{eqnarray}\label{fluc}
 \underline{m} \equiv (m-\Bar{m})~,~\underline{\mu} \equiv (\mu-\Bar{\mu})~,~ \underline{\eta} \equiv (\eta-\Bar{\eta})
\end{eqnarray}
Eqn.s (\ref{sadpt}), (\ref{kment})(\ref{bek}) (\ref{repara})and (\ref{fluc}) together give
\begin{eqnarray}\label{par}
 Z_{G} &=& \exp[ X(\Bar{m},\Bar{\mu},\Bar{\eta})] 
\times  \int d\underline{m}~ d\underline{\mu}~ d\underline{\eta}~ \exp \{\frac{1}{2}[( X_{mm} )\underline{m}^{2} + ( X_{\mu\mu})\underline{\mu}^{2}+ X_{\eta\eta})\underline{\eta}^{2}+(2  X_{m\mu})\underline {m} \underline{\mu}) \nonumber \\
&+& (2  X_{m\eta})\underline {m} \underline{\eta})+((2  X_{\eta\mu})\underline {\eta} \underline{\mu}) \} 
\end{eqnarray}
Where,
\begin{eqnarray}\label{x}
X(\Bar{m},\Bar{\mu},\Bar{\eta}) &=& \log(T)+S-\beta M +\beta \Psi Q + \beta \Omega J \nonumber \\
 &=& \log(m)+\frac{1}{2} \log \cosh \eta-\frac{1}{2} \log (1+ \sin \mu ) + \log(3\cosh^{2} \eta-1)\nonumber \\ 
 &+&2\pi m^{2} \cosh \eta (1+ \sin \mu)-\beta m\Big(1+\frac{\sinh^{2} \eta}{2}\Big)+ \beta \Psi \frac{m}{2} \sinh 2\eta \nonumber \\
 &+&\beta \Omega m^{2} \cos \mu \cosh \eta
 \end{eqnarray}
Here, $T$ is Jacobian due to change of variables from $(A,Q,J)$ to $(m,\mu,\eta)$ and is given as,
\begin{eqnarray}\label{jacobian}
T=8\pi m^{4}(1+ \sin \mu)(3\cosh^{2} \eta-1) \cosh^{2}\eta
\end{eqnarray}
Saddle point approximation method implies $X_{m}=0=X_{\mu}=X_{\eta}$ at the saddle point and hence in the large area limit $X_{\mu}=0$ and eqn. \ref{x} together give 
\begin{eqnarray}\label{1st deri}
\beta \Omega= \frac{2\pi}{\tan \mu}
\end{eqnarray}
In large area limit, eqn.s {\ref{x}} and {\ref{1st deri}} give $X_{mm}=4\pi \cosh \eta \hspace{.03 in} (1+ cosec \hspace{.025 in}\mu)$ and is non-negative. This implies that $Z_{G}$ will diverge and  macroscopically large dilatonic black holes in asymptotically flat spacetime must be thermally {\it unstable} with respect to Hawking radiation, over their entire parameter space.\\

\end{document}